\documentclass{article}

\usepackage{graphicx}
\usepackage{dcolumn}
\usepackage{bm}
\usepackage{hyperref}
\usepackage[utf8]{inputenc}
\usepackage[T1]{fontenc}
\usepackage{enumerate}
\usepackage{color}

\begin{document}
\oddsidemargin= 20mm
\paperwidth= 210mm
\textwidth= 160mm
\topmargin= -10mm
\textheight= 220mm
\marginparwidth= 0mm

\title{McStas (i): \\ Introduction, use, and basic principles for ray-tracing simulations}


\author{P. K. Willendrup, \\ Department of Physics, Technical University of Denmark, Denmark, \\ Data Management and Software Center, European Spallation Source, Denmark, \\ K. Lefmann, \\ Niels Bohr Institute, University of Copenhagen, Denmark}



\date{\today}

\maketitle

\begin{abstract}
We present an overview of, and an introduction to, the general-purpose neutron simulation package McStas. We present the basic principles behind Monte Carlo ray-tracing simulations of neutrons performed in the package and present a few simple examples. We present the implementation of McStas, the status of the package and its use in the neutron community. Finally, we briefly discuss the planned development of the package. 
\end{abstract}


\section{Introduction}

Over the last decades, Monte Carlo ray-tracing simulation has become increasingly important for
investigating the performance of neutron scattering instrumentation. Few neutron instruments are being built before undergoing a thorough investigation by simulation.
In particular, the powerful spallation neutron sources in North America (SNS) \cite{SNS} and Japan (J-PARC MLF) \cite{JPARC}, the upgrade of the UK source (ISIS TS-2) \cite{ISIS}, and the coming European Spallation Source (ESS) \cite{ESS}
have heavily utilized neutron simulation software, thereby also pushing the development of the simulation methods.

At the end of last century, neutron instrumentation ray-tracing
simulations were performed using home-made monolithic codes. While
fairly successful, this approach suffered from limited development ressources -
with potential problems of quality assurance - while still creating a
vast amount of duplicated effort between different projects and facilities. 
One exception from the monolithic approach was the general-purpose neutron
simulation package NISP \cite{NISP,NISPpage} released 1994/95, based
on the earlier \verb+MCLIB+ subroutine library from 1978/1980 \cite{MCLIB}. However, NISP never reached a critical mass of user community worldwide. 

Almost simultanously, at the turn of the century, four general simulation packages were launched, following the NISP philosophy, but on more modern software platforms:
\begin{itemize}
\item The HZB-based VITESS \cite{VITESS}
\item The \v{R}e\v{z}-based RESTRAX/SIMRES \cite{RESTRAX}
\item The SNS-based IDEAS \cite{IDEAS}
\item The Ris\o - ILL based McStas \cite{NN99}
\end{itemize}
Most of these packages are still maintained (to a higher or lower degree) and the status of their projects can be found on their respective homepages \cite{VITESSpage,RESTRAXpage,McStaspage}.

In the view of the authors of this paper, the state of the art of currently available neutron Monte Carlo ray-tracing packages are summarized in the table below, including points of view on  differences and main advantages and disadvantages.
\\\ \\
{\small
\begin{tabular}{|c|l|l|}
\hline {\bf Package}  & {\bf Plusses} & {\bf Minusses} \\ 
\hline VITESS & 
\begin{minipage}[t]{0.4\textwidth}
     \begin{itemize}
         \item Easy for beginners - no code to write.
         \item Use of pipes gives automatically parallel execution. 
         \item Is on track to become hosted at FZJ.
     \end{itemize} 
\end{minipage}
& 
\begin{minipage}[t]{0.4\textwidth}
     \begin{itemize}
         \item Modules historically written without common structure requirement / standard.
         \item Hard to modify at the physics / module level as a user.
         \item Open source, but without open software repository.
         \item Key staff has at periods had limited backing from hosting institution.
     \end{itemize} 
\end{minipage} 
\\ 
\hline 
\begin{tabular}{c}
    RESTRAX \\ SIMRES 
\end{tabular} & 
\begin{minipage}[t]{0.4\textwidth}
     \begin{itemize}
         \item Has  mode for "upstream" simulations from sample to source. Gives a clear speed boost for sparse phase-space simulations.
         \item Historically specialized for TAS.
     \end{itemize} 
\end{minipage} 
& 
\begin{minipage}[t]{0.4\textwidth}
     \begin{itemize}
         \item Historically specialized for TAS.
         \item Open source, but without open software repository.
         \item Only few developers
     \end{itemize} 
\end{minipage} 
\\ 
\hline NISP & 
\begin{minipage}[t]{0.4\textwidth}
     \begin{itemize}
         \item No assumption that instruments should be linear. Any volume can scatter to any volume.
         \item Monte Carlo engine derived from early MCNP. 
     \end{itemize} 
\end{minipage} 
& 
\begin{minipage}[t]{0.4\textwidth}
     \begin{itemize}
         \item No assumption that instruments should be linear. (Intrinsically demanding / slow.)
         \item Monte Carlo engine derived from early MCNP.
         \item Legacy Fortran program with unspecified license terms.
         \item Only few developers
     \end{itemize} 
\end{minipage} 
\\ 
\hline McVine & 
\begin{minipage}[t]{0.4\textwidth}
     \begin{itemize}
         \item Automatically enherits much functionality from McStas.
         \item Instrument description written in pure Python. (Industry standard.)
         \item Interfaces with physics models written in Python/\texttt{c++}
         \item Object-oriented with separated physics/geometry
         \item Open source (BSD) with open repository.
     \end{itemize}
\end{minipage} 
& 
\begin{minipage}[t]{0.4\textwidth}
     \begin{itemize}
         \item Instrument description written in pure Python. (Not a dedicated solution for neutron scatterers.) 
         \item Interfaces with physics models written in Python/\texttt{c++}
         \item Due to object-oriented nature, components only interact via the neutron
         \item Not widely used outside US neutron scattering.
         \item Only few developers.       
     \end{itemize} 
\end{minipage} 
\\ 
\hline RAMP & 
\begin{minipage}[t]{0.4\textwidth}
     \begin{itemize}
         \item New kid on the block.
         \item Intrinsically GPU-capable.
         \item Written in pure Python. 
         \item Open Source with open repository.
      \end{itemize} 
\end{minipage}  
& 
\begin{minipage}[t]{0.4\textwidth}
     \begin{itemize}
         \item New kid on the block, i.e. short history so far.
         \item Functionality so far limited.
         \item Order of magnitude GPU speedup is not yet achieved.
     \end{itemize} 
\end{minipage}
\\ 
\hline McStas 
& 
\begin{minipage}[t]{0.4\textwidth}
     \begin{itemize}
         \item Flexible, users can interact at instrument, component and c-code level.
         \item True Open Source from the beginning, has open repository, many user contributions.
         \item New functionality can be added through short, modular c-code.
         \item Several developers working in parallel for almost full duration of project history.
         \item Parallelised using industry-standard MPI solution. (GPU developments are under way.) 
         \item Interfaces with many other codes e.g. MCNP, Mantid, iFit, 
     \end{itemize} 
\end{minipage}
& 
\begin{minipage}[t]{0.4\textwidth}
     \begin{itemize}
          \item Extremely flexible, users can interact at instrument, component and c-code level. (Which may be error prone if many advanced features are combined.)
          \item True Open Source from the beginning, many user contributions. (There are however requirements for code inclusion.)
     \end{itemize} 
\end{minipage}
\\ 
\hline
\end{tabular}
}

Based on the need for instrument developments, the early use of the software packages concentrated on simulating neutron optics, especially guide systems, and concepts for full instruments, including instrument resolutions. 

A huge effort was performed in the beginning of this century with systematic comparison between simulation results of guides and optics in different packages and to some extent with actual measured data \cite{comparisons}. This work served strongly to track down tricky numerical errors, to increase confidence in the packages themselves, and to establish the ray-tracing technique as such.

With the aim of being able to simulate complete virtual 
experiments, the last decade has seen a growing demand for - and development of - a suite of scattering sample components. 

This is the first article in a series of short reviews of the McStas package - with the planned future articles described in section~\ref{sect:reviewseries}. For this reason, we will here review the general expressions used for Monte Carlo ray-tracing simulations of neutron scattering, In addition, we will discuss the implementation of the McStas system, its utilization and user community, and the planned future development of the package.

We have aimed this article for neutron scatterers and software developers alike, and we thus begin
with a few introductory paragraphs on the nature of neutrons and on the ray-tracing technique. Most material presented here, and other related information, can be found also in the user and component manual for McStas \cite{McStasmanual}.


\section{The mathematics of Monte Carlo ray-tracing simulations for slow neutrons} \label{sect:math}

Monte Carlo simulations are in general a way to perform approximate
solutions to complex problems by use of random sampling, thereby performing \textit{numerical experiments}. The first known example of such a method is that of the \textit{Buffon needle problem}\cite{BuffonBook} first posed in the 18th century by Georges-Louis Leclerc, Comte de Buffon:

{\hspace{0pt} \\ \textit{"Suppose we have a floor made of parallel strips of wood, each the same width, and we drop a needle onto the floor. What is the probability that the needle will lie across a line between two strips?"}\hspace{0pt} \\}

As explained in a wikipedia article\cite{Buffon}, the thinking behind that numerical experiment can in fact be used to construct a Monte Carlo algorithm for the estimation of $\pi$. 
In terms of an actual mathematical algorithm, and later code for computers, the first formulation was in fact to facilitate calculations of neutron physics for the Manhattan project \cite{metropolis}.

As with all stochastic methods, Monte Carlo methods are prone to 
statistical variances. To reduce the statistical errors, a number of {\em variance reduction} methods have been designed, of which we here mention two. In
{\em Stratified sampling}, one divides the parameter space up into mutually exclusive strata, and sample a given amount of each stratum.
In {\em Importance sampling}, one will choose to sample more often in regions more important for the final result.
As an analogy, to calculate the average height over sea level of a landscape, one would map the ragged mountain slopes more carefully than the still water in the mountain lake. 

The method relevant for neutron simulations is known as {\em Monte Carlo ray-tracing}. This is used to study objects which travel along a path (a ray), and can be (partially) absorbed or scattered into another direction, but not converted into other types of radiation.
The most known example of this is visible light, and this type of
ray-tracing is frequently used to generate realistic illumination in
3D computer graphics, see e.g. Pov-Ray \cite{povray}.

\subsection{The semiclassical representation of neutron coordinates}
In neutron ray-tracing simulations a neutron is represented semiclassically 
by simultaneously well-defined position, {\bf r}, velocity, {\bf v}, time, $t$, and all three components of the neutron spin vector, {\bf s}. 

Formally, this approach violates the laws of quantum mechanics, in particular
the Heisenberg uncertainty relations \cite{merzbacher}, 
given for pairs of conjugate position/momentum variables by 
\begin{equation}
\delta E \delta t \geq \frac{\hbar}{2} \; ; \; \delta x \delta p_x \geq \frac{\hbar}{2} ,
\end{equation}
and similarly for the conjugate variables in the $y$- and $z$-directions.
However, the semiclassical approximation works really well for describing instruments that use ``typical'' neutrons with velocities of the order $v \approx 100 - 10000$~m/s. For example we can imagine a neutron beam with the typical value $\lambda = 4$~\AA\ ($v=1000$~m/s) and a low divergence of $0.06^\circ$ (0.001~radians) passing a tight slit of 0.1~mm width. Here we have $\delta x = 10^{-4}$~m and  $\delta p_x = m_{\rm n} \delta v_x = 0.001 \, m_{\rm n} v_x \approx 1.7 \cdot 10^{-26}$~kg m/s. Hence, the product of the two uncertainties is around $1.7 \cdot 10^{-30}$~kg m$^2$/s, or 16000 $\hbar$, so the uncertainty relation holds with a large margin. As another example, the best energy resolution of a backscattering instrument is around $\delta E \approx 0.5\, \mu$eV for neutrons of 6.6~\AA , and the best time resolution is the time is takes one of these neutrons to transverse a small (2~mm) sample, $\delta t \approx 3\, \mu$s. The product of this is $1.5 \cdot 10^{-12}$~eVs, or 400 $\hbar$. Hence, the uncertainty relation also here holds well. 

The validity of the semiclassical approach for neutron scattering is also discussed in detail in Ref.~\cite{mezei99}. The similar issues with the semiclassical approximation of the neutron spin will be discussed in the future review on polarized scattering.\\

\subsection{The weight factor}
\begin{figure}
\includegraphics[width=0.7\columnwidth]{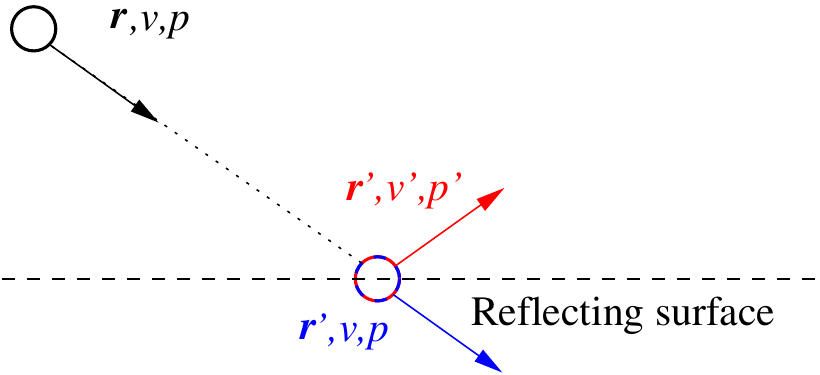}
\caption{Neutron ray interacting with a reflecting surface: 1) The
  neutron begins with parameters {\textit {\textbf r},v,p} (black) at a distance from
  the surface. 2. The neutron is propagated to the surface and now has
  parameters {\textit {\textbf r}',v,p} {\color{blue}(blue)}. 3. The neutron is reflected and
  achieves updated parameters {\textit {\textbf r}',v',p'} {\color{red}(red)}.} \label{fig:state}
\end{figure}

The physical unit of a neutron ray is neutron rate, or neutron intensity (s$^{-1}$). On a typical neutron source, the emission rate can amount to $10^{15}$/s; an order of magnitude larger than the number of rays that can typically be simulated on a laptop in one year.
For this reason, to simulate realistic values for neutron count numbers, 
a neutron ray in general represents more than a single physical neutron (per second). 
To accommodate this, each ray contains an additional parameter,
the {\em weight factor}, $p$, with a unit of neutrons per second. When the ray begins at the source, $p$ is typically $10^3$/s to $10^9$/s. The intensity of the given neutron ray is updated on interaction with components of the neutron instrument, see Figure \ref{fig:state} for an illustration.

As one example of importance sampling, the weight factor is manipulated through the simulation. For example, when some physical neutrons are ``lost''
due to {\em e.g.} finite reflectivity or absorption, the simulated ray
will in general continue in the simulations, 
while $p$ is adjusted to reflect the correct average physical behaviour. 
When (if) the neutron ray reaches the detector,
$p$ may in practice be less than 1/s. Turning again to Figure \ref{fig:state}, using importance sampling by describing the reflected beam only will not neglect the physical influence of absorption, since we ensure conservation of intensity, i.e. $p=p'+p''$, where $p''$ is the intensity of the non-reflected beam (which we in this example do not transport).

We proceed along this line of reasoning to calculate the neutron intensity in a beam. Consider a neutron component at a surface perpendicular to the beam. The neutron intensity is given by the sum of all simulated rays reaching this surface: 
\begin{equation}
I_j = \sum_{i=1}^N p_{i,j-1},
\end{equation}
where $i$ is the ray index, $N$ is the total number of rays, 
and $j$ is the index of the given component. The index $j-1$ on $p$ indicates that we consider
the intensity being emitted by the previous component. 
If the ray does not reach this point, we have
$p_{i,j-1}=0$. 
The weight of the neutrons after interacting with component $j$, is expressed by
\begin{equation}
p_{j} = w_{j} p_{j-1} ,
\end{equation}
where the ray index, $i$, is omitted for simplicity.
The {\em weight multiplier} of the $j$'th component, 
$w_j$, is calculated by the important probability rule
\begin{equation} \label{eq:weight_master}
f_{\rm MC, b} w_j = P_{\rm b}
\, ,
\end{equation}
where $P_{\rm b}$ is the physical probability for event b, and 
$f_{\rm MC, b}$ is the Monte Carlo sampling probability.

Often, there is only one non-trivial event to consider.
This may, {\em e.g.}, be the case for a neutron beam being attenuated by absorption.
Here, each neutron ray either survives or is lost. Absorbed neutrons are typically not simulated ($f_{\rm MC}=0$), while for the other case (transmission) $f_{\rm MC}=1$. The probability rule thus dictates that $w_j=P_{\rm b}$.

When a Monte Carlo branch point is reached (selection between several events), we have $f_{\rm MC, b} < 1$ for each branch, b. 
However, since $f_{\rm MC}$ is a probability function, we must have
\begin{equation}
\sum_b f_{\rm MC, b} = 1 .
\end{equation}
If a variance reduction scheme is used where some branches are simulated less frequently than the physical probability,
we have $w_j = P_{\rm b} / f_{\rm MC, b}$, which may exceed unity. Hence, the weight factor may also increase during the simulation.

\subsection{Neutron moderators and focusing}
Consider a component that may send neutrons out in all directions, {\em e.g.}\ the moderator or the sample. Often in McStas, we choose uniform sampling 
when simulating all neutron directions. The sampling probability per solid angle is then:
\begin{equation}
f_{\rm MC\, unfocus} d\Omega = \frac{d\Omega}{4 \pi} .
\end{equation}
However, this simulation scheme may cause huge inefficiencies. In many simulations, only rays emitted in certain ``interesting directions'' have any chance of being detected. 
In such cases, one will employ the technique of {\em focusing},
where the simulated ray will be emitted only within a certain solid angle, $\Delta \Omega$ - but with uniform probability within $\Delta \Omega$. 
Then the Monte Carlo sampling probability is:
\begin{equation}
f_{\rm MC\, focus} d\Omega = \frac{d\Omega}{\Delta \Omega} .
\end{equation}
To avoid systematic errors in the simulation results, $\Delta \Omega$ should contain all the directions contributing to the neutron intensity to be simulated.
When using (\ref{eq:weight_master}) to obtain the weight factor transformations, we see that the focusing alone will bring the amount
\begin{equation}
w_{\rm focus} = w_{\rm unfocus} \frac{\Delta\Omega}{4\pi} 
 \, ,
\end{equation}
which is the general focusing contribution to the weight factor. Comparing to the uniform case, the focusing method gives smaller weight 
factors per ray, but a larger number of rays traveling towards the detector.
This gives, on average, the same final result, but with a smaller statistical error.
The focusing technique is a typical example of importance sampling.

Imagine a McStas moderator component, simulating a moderator that emits $R$ neutrons per second, divided into $N$ rays. The total intensity in the simulation is found from
\begin{equation}
R = I_0 = \sum_{i=1}^N p_{i,0} .
\end{equation}
Assuming uniform distribution of ray weights, we find the initial weight to be $p_0 = R/N$. If focusing of the emitted rays is taken into account, we reach the expression for the initial weight
\begin{equation}
p_0 = \frac{R}{N} \frac{\Delta \Omega}{4 \pi} . 
\end{equation}
Most moderators have wavelength-dependent emission rates, where $R = \int \phi(\lambda) d\lambda$. Imagine that we sample uniformly a wavelength band of width $\Delta \lambda$. Then the sampling frequency would be $f_\lambda = 1/\Delta \lambda$. With the same argument as the focusing, we reach the final expression for the moderator weight factors
\begin{equation}
p_0(\lambda) = \frac{\phi(\lambda)}{N} \frac{\Delta \Omega}{4 \pi} \Delta \lambda,
\end{equation}
which is the typical equation being used in McStas moderator components.

\subsection{Estimates of simulation uncertainty}
In a simulation, like for real experiments, it is important to be able to estimate the statistical uncertainty. We here present a simple derivation of the uncertainty in simulations with weight factors.

Imagine $M$ rays, all with weight factor $p$. Each 
ray is imagined to have an overall probability, 
$P_{\rm d}$, of reaching the detector. The number of observed rays will be binominally distributed with a mean value $\bar{N} = MP_{\rm d}$ and variance 
$\sigma^2(N) = M P_{\rm d} (1-P_{\rm d})$ \cite{barlow}. 
Very often, we will have $P_{\rm d} \ll 1$, 
but still sufficiently large values of $M$, so that
$\bar{N} = M P_{\rm d} \gg 1$. In this case,
the distribution will be approximately Gaussian with a variance $MP_{\rm d}$ - or a standard deviation
\begin{equation}
\sigma(N) = \sqrt{MP_{\rm d}} \approx \sqrt{N} ,
\end{equation}
where the observed value, $N$, is used as the best estimate for the true mean value, $\bar{N}$.
Recalling the constant weight factors of the rays, the total simulated intensity becomes 
\begin{equation} 
I=Np, 
\end{equation}
with a variance 
\begin{equation} \label{eq:MCerror1}
\sigma^2(I) = \sigma^2(N) p^2 = Np^2 .
\end{equation}

We now allow the rays to have different, but discrete, weight factors, $p_i$.
The simulated number of rays with one particular weight factor is denoted 
$n_i$ (standard deviation $\sqrt{n_i}$). The total simulated result is now
\begin{equation}
I = \sum_i n_i p_i .
\end{equation}
Using the property of uncorrelated events, $\sigma^2(a+b) = \sigma^2(a) + \sigma^2(b)$, we can rewrite the sum to reach the statistical variance of the simulated result:
\begin{equation} \label{eq:MCerror2}
\sigma^2(I) = \sum_i n_i p_i^2 . 
\end{equation}

Now, let the simulations occur with rays of arbitrary weight 
factors, $p_j$, different 
for each ray, {\em i.e.} $n_i \equiv 1$, in practice allowing a continuous distribution of $p_i$. 
If the distribution of these weight factors is reasonably well behaved, we can generalize the equations above
to reach
\begin{equation}
I = \sum_i p_i 
\, ,
\end{equation}
with a variance:
\begin{equation} \label{eq:MCerror_final}
\sigma^2(I) = \sum_i p_i^2 
\, ,
\end{equation}
which is, in the proper limits, consistent with (\ref{eq:MCerror1}) and (\ref{eq:MCerror2}).
This is the principle for calculating uncertainties of simulation results in the McStas package. In practice, the sum of both $p_i$ and $p_i^2$ are being accumulated during the simulation to avoid having to store the full list of $p_i$'s for rays reaching the detectors.

\subsection{Scaling to real-world statistics}
In order to compare with real experiments, in particular when using simulated data as input to data analysis programs, simulated intensities must be converted to average integer counts using an imagined counting time, $T$. However, this time cannot be chosen to be arbitrarily high, since the data analysis program will assume $\sqrt{N}$ statistics, and for high values of $T$, the counting statistics will be better than that of the simulation, causing unreal errors in the data treatment. 

Let us now we quantify the effect of scaling with a counting time. 
The average count number is easily obtained by 
\begin{equation}
\left\langle N \right\rangle = I T,
\end{equation}
with $I$ being the simulated intensity. Now, care must be taken when evaluating the standard deviations. Counting statistics require that the variance is 
\begin{equation}
\sigma^2(N) = \langle N \rangle ,
\end{equation}
while the simulated value will have the simulated variance $\sigma^2(IT) = \sigma^2(I) T^2$. We must require that the percieved counting statistics cannot be better than the simulated variance, or $N \geq \sigma^2(I) T^2$, leading to
\begin{equation} \label{eq:timemax}
T^2 \leq \frac{\langle N \rangle}{\sigma^2(I)}. 
\end{equation}
For many virtual experiments, the value of $T$ can be surprisingly low, of the order seconds, since in particular the highest counting rates tend to be underrepresented in the simulations. 

If a series of simulated values need to be scaled by the same time, $T$, the maximum time is given by the smallest value of $T$ that fulfills (\ref{eq:timemax}), or 
\begin{equation} 
T_{\max}^2 = {\rm Min}_j \left\{ \frac{\langle N_j \rangle}{\sigma^2(I_j)} \right\}. 
\end{equation}
Now, when performing a scaling of this type, some data points will likely obtain actual errors smaller than counting statistics.
To make sure that every point in the series has an errorbar that complies with $\sqrt{N}$, one will for all points, except the one with the smallest value of $\langle N_j \rangle/(\sigma^2(I_j))$, need to add a random error, $\Sigma_j^2$:
\begin{equation}
\Sigma_j^2 = \langle N_j \rangle - \sigma^2(I_j) T_{\rm max}^2 ,
\end{equation}
and then round up or down to an integer.

In conclusion, the count-error-true simulation numbers are
\begin{equation}
N_j = {\rm Round}\left( I_j T_{\rm max} + E_j\right) ,
\end{equation}
where $E_j$ is a normally distributed stochastic variable with mean zero and variance $\Sigma_j^2$.

This transformation to integer count numbers is not used in McStas, but must be applied as a post processing of the data after the simulations.
Further work is, however, needed on the sampling strategies for future more efficient transformation of errorbars.

\section{The concept and implementation of McStas}

As also outlined in the introduction, the philosophy and mindset
behind starting the McStas project, as laid out by K.~N.~Clausen, was to a large extent to minimize
duplication of code and efforts by allowing code to be shared \cite{NN99}. The main ideas that were implemented since the first version of McStas were:
\begin{itemize}
  \item Licensing should follow Open Source standards. Since version 1.8 McStas is
      released under the GNU General Public License v. 2.0. Earlier
      versions were licensed under a McStas- and RIS\O\ specific
      license that was open, but did not allow redestribution.
  \item The code should be modular to allowed sharing models of instrument parts between users. This evolved into the so-called components, explained below.
  \item The syntax of the user-supplied instrument definition should
      be simple and powerful. For the development of the syntax details, K.~Nielsen got input to the syntax, using pen and paper before writing any code, from the RIS\O\ staff, 
      in particular K. Lefmann and H. M. R\o nnow wrote the first instrument and handful of components in this way.
    \item The software concepts should first of all make sense to
      instrument scientists, allowing them to work in a way that felt
      almost \emph{like runnning an actual instrument}
    \item Documentation should as far as possible be \emph{embedded}
      within the code
    \item Simplicity and readability of the code should be preferred over performance.
\end{itemize}

McStas is implemented in a three-layered structure. The {\bf core system and run-time libraries} take care of the compiling and execution of the program, user interface, and visualization of the results. 
The core and run-time c-code libraries typically come in a set of header and c-code files, such as
\begin{itemize}
    \item \verb+mccode-r.h+ and \verb+mccode-r.c+, common to both McStas and McXtrace and containing functions for generating random numbers, routines for intersecting particle ray and geometrical objects etc.
    \item \verb+mcstas-r.h+ and \verb+mcstas-r.c+, containing physical constants for neutron physics, propagation routines with / without gravitation etc. (McXtrace has a similar set of files implementing X-ray physics)
    \item \verb+interoff-lib.h+ and \verb+interoff-lib.c+ for handling of surface-polygon based geometries in \\ GeomView \verb+OFF+ file format\cite{GeomView}
    \item \verb+interpolation-lib.h+ and \verb+interpolation-lib.c+, routines for interpolating in gridded and sparse datasets
    \item \verb+ref-lib.h+ and \verb+ref-lib.c+, routines for describing the physics of reflective super-mirror surfaces
    \item \verb+read_table-lib.h+ and \verb+read_table-lib.c+, routines for reading tabulated data from files
    \item \verb+pol-lib.h+ and \verb+pol-lib.c+, routines for handling polarized neutron transport
    \item \verb+adapt_tree-lib.h+ and \verb+adapt_tree-lib.c+, routines to define adaptive importance sampling schemes
\end{itemize}

These levels of code are only modified by a handful of system specialists. The core system also manages the main Monte-Carlo loop that in turn produces a specified number of neutron rays for the simulation.
\\ \\ 
Supplementing the core system run-time codes, a number of c-files are present to support individual McStas components, such as
\begin{itemize}
    \item \verb+monitor_nd-lib.h+ and \verb+monitor_nd-lib.c+, supporting the very general Monitor\_nD monitor component from Emmanuel Farhi, ILL
    \item \verb+ESS_butterfly-lib.h+, \verb+ESS_butterfly-geometry.c+ and \verb+ESS_butterfly-lib.c+, supporting the ESS\_butterfly source component
    \item \verb+Geometry_functions.c+, \verb+Union_functions.c+ and \verb+Union_initialization.c+, supporting the Union component set by Mads Bertelsen, ESS DMSC
    \item \verb+nxs.h+ and \verb+sginfo.h+, supporting the Sample\_nxs component from Mirko Boin, HZB
    \item \verb+ess_source-lib.h+ and \verb+ess_source-lib.c+, supporting the legacy ESS\_moderator source component
    \item \verb+chopper_fermi.h+, \verb+chopper_fermi.c+, \verb+general.h+, \verb+general.c+, \verb+intersection.h+,\\ \verb+intersection.c+, \verb+vitess-lib.h+ and \verb+vitess-lib.h+, supporting the Vitess\_Chopperfermi component, ported from VitESS
\end{itemize}

The {\bf components} are modular pieces of code. The McStas components are written in a unique domain-specific language, which is augmented using ANSI-C. A McStas component will typically be equivalent to a beam-optical component, such as a chopper, a collimator, a particular sample, or a detector (often called ``monitor'' in McStas). Sources and moderator are in McStas implemented as one combined component. Components are typically parametrised, \emph{e.g.} with respect to size, in order to be of general use. The components will use the matematical tools for \emph{e.g.}\ weight transformation and focusing, presented in section~\ref{sect:math}.

\begin{figure}
\includegraphics[width=0.7\columnwidth]{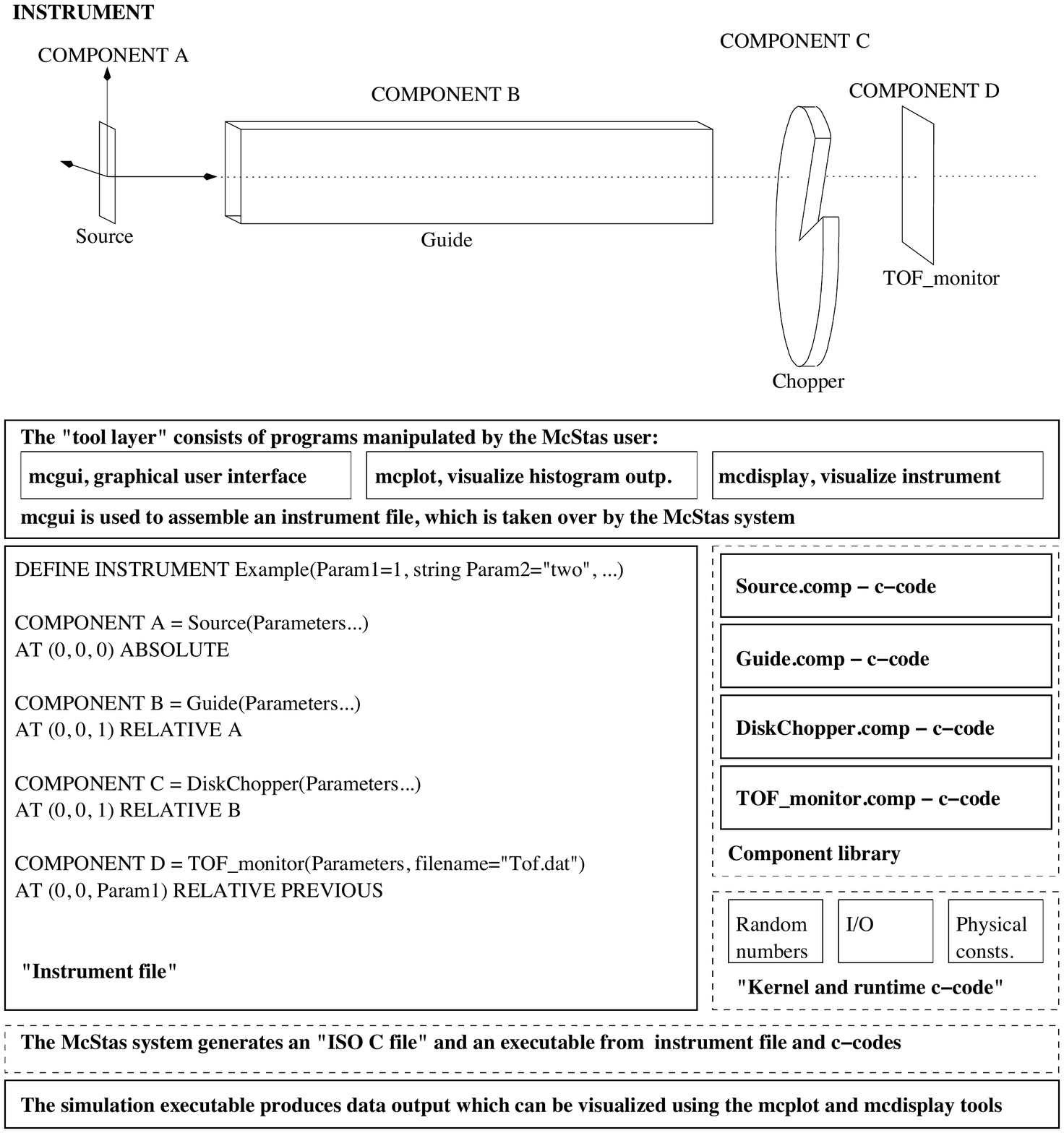}
\caption{The figure illustrates how the McStas (and McXtrace) codes
  are layered, including how user, developers and code generation
  mechanism interact with the various layers of code. The geometrical lay-out of the instrument to simulate is shown at the top panel, while the bottom panes represent the different interacting layers of McStas. The instrument file is shown fully in the large panel to the left; the figure is not to scale. Reproduced from \cite{NOP2014}} \label{fig:mc_structure}
\end{figure}

The {\bf instrument} is a description of a particular neutron scattering set-up. McStas operates at one instrument at a time. The instrument file contains a selection of components and their spatial position and orientation. The instrument is written fully in the McStas domain specific language, although ANSI-C may be used to augment the functionality. Also the instruments can be parametrised, for example to allow the scanning of a triple-axis instrument by changing a number of angles. 
Fig.~\ref{fig:mc_structure} illustrates the roles of system, components, and instrument file for a very simple time-of-flight instrument, consisting of a pulsed source, a guide, a chopper, and a time-of-flight detector. 
 
When McStas is executed on an instrument file, the system compiles the information from the instrument file and the component library and produces an ANSI-C file. This is in turn compiled by a standard C-compiler to an executable code, which is called by the system. After the process has terminated, McStas collects the resulting data files, stores them in specified directories, and visualises the data upon request. Everything can be controlled from a user-friendly GUI interface, or by a scripting language.

McStas has a users manual and a component manual \cite{McStasmanual}, both of which are very comprehensive documents. A more readily accessible documentation may be found in the project home page. Documentation of component functionality is furthermore embedded in the header of the component code. This documentation is \emph{e.g.}\ used for the generation of online component help at the McStas home page \cite{McStaspage}.

\section{Use and users of McStas}

In its first two decades, McStas has seen a steady increase of users. We have chosen to illustrate this with a literature search. We have analyzed the 350 articles citing one (or more) of the basic McStas release articles, and by far most of these articles in fact represent simulation work done by the use of McStas. 

\begin{figure}
    \centering
    \includegraphics[width=0.6\columnwidth]{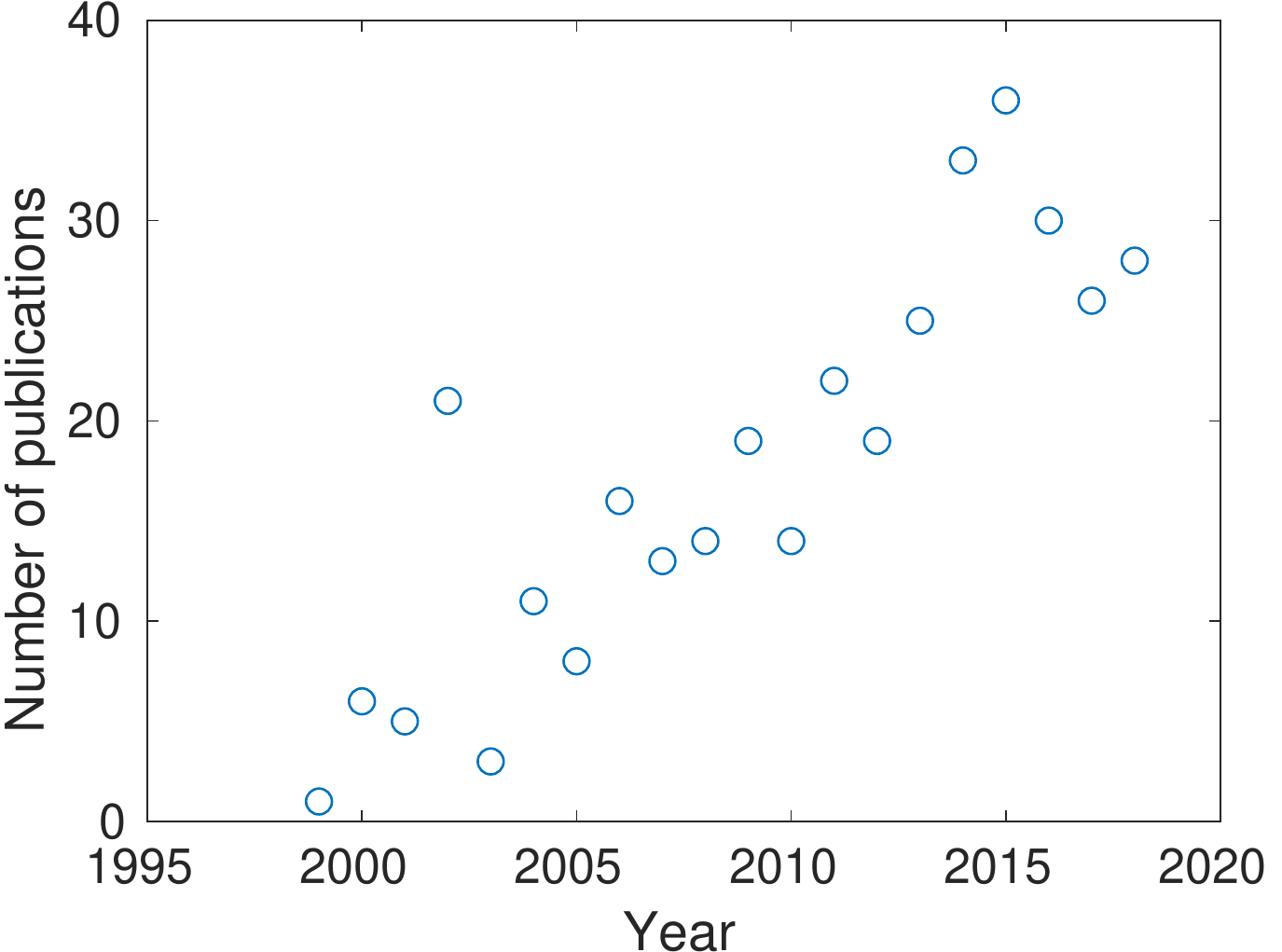}
    \caption{The annual number of articles citing McStas shown as a function of publication year. }
    \label{fig:uses}
\end{figure}

In Fig.~\ref{fig:uses}, we show the number of McStas-using articles as a function of publication year. We see that after a slow start (except for a spike in 2002), the publication rate reached 10/y in 2006 and 20/y in 2011. Since 2014, the number has been approx.\ 30/y, which is partly based on the large instrumentation work related to ESS. 

%
%
%

\begin{figure}
    \centering
    \includegraphics[width=0.6\columnwidth]{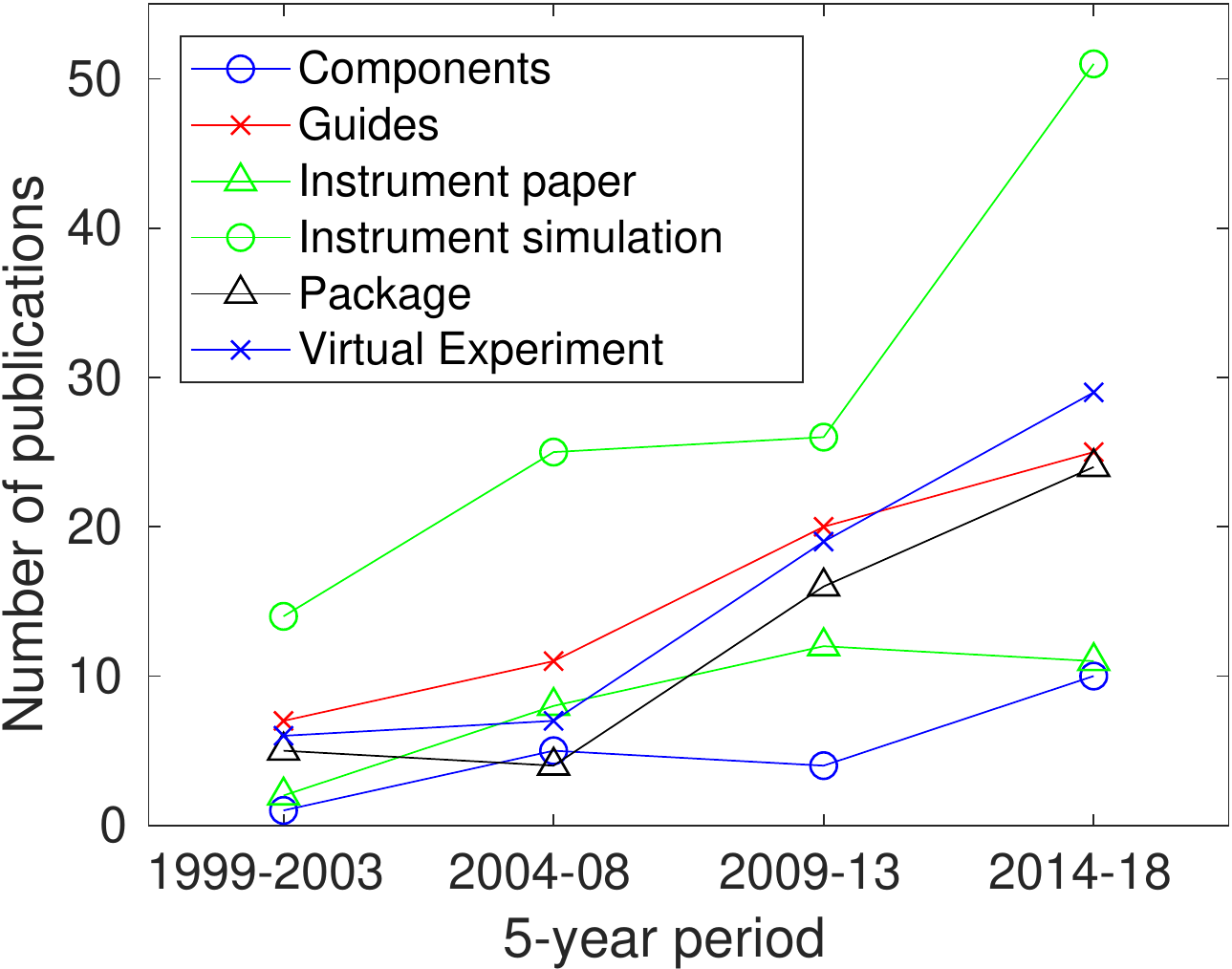}
    \caption{The number of articles citing McStas, divided into use categories, shown as a function of 5-year intervals.}
    \label{fig:use_categories}
\end{figure}

It is highly interesting to study what McStas has been used for, judging from this literature base. Fig.~\ref{fig:use_categories} shows the six main use categories of the packages, distributed on 5-year periods. Although the overall use of McStas has increased by a factor 4 over the years, the distribution between the categories has remained remarkably constant. Let us present the themes one by one (all will be described in much more detail in later issues of this article series):

\paragraph{Components} These papers describe the simulations of individual beam-optical components, such as choppers, monochromators \cite{santisteban05}, lenses \cite{frielinghaus09}, or mirrors \cite{stahn12}. Around 5\% of the articles fall in this category.

\paragraph{Guides} A particular application of McStas has been to study different guide systems, both generic guide geometries \cite{schanzer04,kleno12}, and guide constructions for actual instruments \cite{ibberson09,Le13}. Recently, tools have become available for semi-automatic optimization of guide geometries \cite{bertelsen17} and guide coatings \cite{olsen19}, strongly enhancing the capabilities and speed of simulation-guided optimizations for guide designs. 
Around 20\% of the McStas articles are in this category.

\paragraph{Instrument papers} This denotes the important papers describing the performance of an actual instrument. Such papers will typically obtain many citations, as they will be the standard reference for users of the instrument. Often, the use of McStas in these papers is related to prediction of flux on sample, and the calculation of resolution functions, both to be compared to the actual measured values. Approx.\ 10\% of the McStas papers are instrument papers and represent instruments at the major facilities around the world, for example: 
\begin{itemize}
    \item SNS ({\em e.g.}\ CNCS and ARCS) \cite{ehlers11,abernathy12}
    \item NIST (MACS) \cite{rodriguez08}, 
    \item J-PARC ({\em e.g.}\ AMATERAS and 4SEASONS \cite{nakajima11,kajimoto11}
    \item CARR (HIPD) \cite{yang18} 
    \item ANSTO (TAIPAN) \cite{danilkin12}
    \item ILL (D7) \cite{stewart09}
    \item ISIS (ENGIN-X) \cite{santisteban06}
    \item FRM-2 (KWS-2) \cite{radulescu12}
    \item HZB (FLEX) \cite{Le13}
    \item PSI (RITA-II) \cite{lefmann06}
\end{itemize}  

\paragraph{Instrument simulations} These articles represent the design work behind instruments, some of which are not yet built, and a few of which may never be built at all. This is a main category for McStas work, containing 35\% of all articles. The simulations cover all facilities listed above, as well as:
\begin{itemize}
\item{the Algier reactor}
\item{the Argentine reactor}
\item{CSNS}
\item{Dhruva}
\item{ESS}
\item{HANARO}
\item{Dubna}
\item{LLB}
\item{PIK}
\end{itemize}
Most of the instrument simulation publications are related to the newest sources: ESS (24 papers), J-PARC (20 papers), and FRM-2 (13 papers), but also much activity is related to ILL (12 papers) and SNS (7 papers). Examples are too plenty to be given at this stage, but we will elaborate on the topic later in this review series.


As a separate task, McStas has been used in two demanding studies of the complete instrument suite for ESS, first to optimize the pulse structure \cite{lefmann13}, then to optimize the moderator geometry \cite{andersen18}.

\paragraph{Virtual experiments} These articles represent the most advanced use of simulations, namely the performance of a full virtual experiment on a simulated instrument \cite{Lefmann08,farhi09}. This has been used for the determination of accurate resolution functions \cite{udby11,eckold14}, tools for teaching practical neutron scattering without the investment of beam time \cite{udby19}, for correcting for multiple scattering events in samples \cite{farhi09,bertelsen18}, and even for analysis of complete data sets \cite{boin12,woracek14,taminato16,jacobsen18}. 
Around 15\% of the articles fall in this category.

\paragraph{Packages} These articles describe simulation packages, typically McStas, but also VITESS and RESTRAX have cited McStas, partially for the reason that these three packages have inspired and cross-fertilized each other over the last two decades. In addition, McStas has spawned an X-ray cousin (McXtrace) \cite{knudsen13} and a He-scattering cousin \cite{eder17}. Around 15\% of the articles are in this category.\\

In addition to this bibliometric study, we have performed a survey of the McStas user community during the autumn of 2018. Ten questions were posed via the SurveyMonkey platform \cite{SurveyMonkey} in the period from November 6th 2018 to December 3rd 2018. All questions were voluntary / optional and contained the possibility of free-form input via an 'other' text field. In total 52 responses were received, out of which 47 filled demographic data. The full McStas user base is estimated to be approximately 400, based on the current 252 members of the McStas user mailinglist and knowing from experience that not all users register.\\\ \\

The main findings of the survey are that: 
\begin{enumerate}[1.]
    \item The McStas users are to a high degree running the most recent version of McStas, at the time of survey v. 2.4.1.
    \item Linux is the dominant operating system, followed by Windows 10 and macOS. Some users are now running the Linux binaries under Windows 10 via the Windows Subsystem for Linux\cite{WSL}.
    \item McStas users are distributed over all of the continents that have access to neutron facilities.
    \item McStas users are mostly based at neutron facilities, but the package also finds use at universities and in industry.
    \item Most users are now routinely using our modernised Python based gui and command-line utitlites.
    \item The vast majority of McStas users are very happy about the McStas installation process, daily use and the available user support.
\end{enumerate}
For those interested in the full details, a full report of the survey is published at the McStas website\cite{SurveyResults}.

\section{The future of McStas}
As our user survey demonstrates, McStas is as a project alive and healthy with a 20 year long history of development and a stable user community. What separates McStas from the alternative simulation packages is mainly the applied code-generation technique, the quality of documentation and the strongly  collaborative nature of the package, allowing users to contribute. To stay alive and healthy however, a sustained effort is needed in terms of manpower, resources, support and development. \\\ \\
The main focal points for the development in the coming years are: 
\begin{enumerate}[1.]
    \item Lower the typical release cycle of 12-18 months to 3-6 months. Updating more often lowers 'time to market' for new ideas, solutions and bug-fixes. McStas 2.5 was released in December 2018 and 2.5.1 will be released in the spring of 2019.
    \item Make our test/validation effort more stringent and use more of a 'continuous integration' approach to testing. All components should have output test data that we can monitor against on a continuous basis. Today only a scalar numerical output defines a 'test' - in the future we would like to compare at a more detailed level of data in individual bins of individual monitors.
    \item Enrich the sample library even further. A natural layer of growth is the Union \cite{bertelsen18} concept by Mads Bertelsen which allows to separate geometry and physical processes by defining an \emph{inner Monte Carlo loop} (e.g. for describing complicated arrangements of matter in / around the sample area). The Union components should become complete in terms of describing all possible geometries and all possible physics, which is today not fully the case.
    \item A modernization of the code generation layer is under way, which will allow us to move to a \texttt{C}-function based grammar rather than today's \texttt{\#define}-based approach\footnote{The classical McStas code uses c-preprocessor\cite{preprocessor} define instructions to locally change the meaning of a variable or parameter name. This means that e.g. a geometrical parameter symbol \texttt{xwidth} can be used by multiple components within the same instrument, but with a local meaning and without inconsistency.}. This will break certain backward-compatibility and move us to McStas v. 3.0
    \item The modernized code-generation will ease the process of targeting modern GPU-based architectures.
\end{enumerate}
The reader is further encouraged to consult our standard references \cite{NOP2014} and \cite{McFuture} for a relatively recent and complete description of McStas.

\section{The McStas review series} \label{sect:reviewseries}
This article should be seen as an introduction to a series of McStas review papers. Planned themes in this series cover the main uses of McStas described above:
\begin{itemize}
    \item Components
    \item Guide systems
    \item Instrument simulations and virtual experiments
\end{itemize}
In addition to a review of these important cases, we also will describe a few more technical matters that deserve a more thorough presentation than what has been given in literature so far:
\begin{itemize}
    \item Modeling of scattering from samples
    \item Simulation of polarized neutrons, including spin precession
\end{itemize}
Is is our aim that this article series on McStas will serve to share knowledge of the package and its utilization, as well as enhance the capabilities for designing and modeling neutron instrumentation worldwide.

\section*{Acknowledgements}
It is a pleasure to thank everyone involved in the McStas project over the decades. In chronological order: K. N. Clausen, K. Nielsen, H. M. R\o nnow, E. Farhi, P.-O. \AA strand, K. Lieutenant, P. Christiansen, E. B. Knudsen, U. Filges, J. Garde, and M. Bertelsen. 

\end{document}